\documentstyle[12pt]{article}

\newcommand{\la}{$\lambda$}
\newcommand{\oiii}{[O\,{\sc iii}]}
\newcommand{\oiiia}{\oiii\,\la4959}
\newcommand{\oiiib}{\oiii\,\la5007}
\newcommand{\feii}{Fe\,{\sc ii}}

\newcommand{\feiib}{\feii\,\la5018}

\newcommand{\lya}{Ly$\alpha$}
\newcommand{\ha}{H$\alpha$}
\newcommand{\hb}{H$\beta$}
\newcommand{\hei}{He\,{\sc i}}
\newcommand{\heia}{\hei\,\la5876}
\newcommand{\heii}{He\,{\sc ii}}
\newcommand{\heiia}{\heii\,\la4686}
\newcommand{\ciii}{C\,{\sc iii}]}
\newcommand{\ciiia}{\ciii\,\la1909}
\newcommand{\civ}{C\,{\sc iv}}
\newcommand{\civa}{\civ\,\la1549}
\newcommand{\fevii}{[Fe\,{\sc vii}]}
\newcommand{\feviia}{\fevii\,\la5721}
\newcommand{\feviib}{\fevii\,\la6086}
\newcommand{\mgii}{Mg\,{\sc ii}}
\newcommand{\mgiia}{\mgii\,\la2798}
\newcommand{\ergscm}{erg~s$^{-1}$~cm$^{-2}$}
\newcommand{\ergscma}{\ergscm~\AA$^{-1}$}
\newcommand{\f}{\footnotemark}
\newcommand{\fn}[1]{\footnotemark[#1]}

\begin{document}


\begin{center}
\bf STEPS TOWARD DETERMINATION OF THE SIZE\\
AND STRUCTURE OF THE BROAD-LINE REGION\\
IN ACTIVE GALACTIC NUCLEI.\\
VI. VARIABILITY OF NGC 3783 FROM GROUND-BASED DATA$^\star$

\bigskip

\sc
G.~M.~Stirpe,\f\
C.~Winge,\f$^,$\f\
B.~Altieri,\f$^,$\f\
D.~Alloin,\f\
E.~L.~Aguero,\f$^,$\f\
G.~C.~Anupama,\f\
R.~Ashley,\f\
R.~Bertram,\f$^,$\f\
J.~H.~Calderon,\fn7\
R.~M.~Catchpole,\f\
R.~L.~M.~Corradi,\fn4$^,$\f\
E.~Covino,\f\
H.~A.~Dottori,\fn2\
M.~W.~Feast,\fn{10}\
K.~K.~Ghosh,\f\
R.~Gil~Hutton,\f$^,$\f\
I.~S.~Glass,\f\
E.~K.~Grebel,\fn4\
L.~Jorda,\fn4$^,$\fn6\
C.~Koen,\fn{19}\
C.~D.~Laney,\fn{19}\
M.~Maia,\f\
F.~Marang,\fn{19}\
Y.~D.~Mayya,\f\
N.~Morrell,\f\
Y.~Nakada,\f\
M.~G.~Pastoriza,\fn2\
A.~K.~Pati,\fn{21}\
D.~Pelat,\fn6\
B.~M.~Peterson,\fn{11}\
T.~P.~Prabhu,\fn{21}\
G.~Roberts,\fn{19}\
R.~Sagar,\fn{21}\
I.~Salamanca,\fn6\
K.~Sekiguchi,\fn{19}\
T.~Storchi-Bergmann,\fn2\
A.~Subramaniam,\fn{21}\
H.~van~Winckel,\fn4$^,$\f\
F.~van~Wyk,\fn{19}\
M.~Villada,\fn7$^,$\fn8\
R.~M.~Wagner,\fn{11}$^,$\fn{12}\
P.~A.~Whitelock,\fn{19}\
H.~Winkler,\f\
J.~Clavel,\fn5\
M.~Dietrich,\f\
W.~Kollatschny,\fn{26}\
P.~T.~O'Brien,\f\
G.~C.~Perola,\f\
M.~C.~Recondo-Gonz\'alez,\f\
P.~Rodriguez-Pascual,\fn{29}\
and
M.~Santos-Lleo\fn6\

\bigskip

\end{center}

{
\footnotesize
\parindent=-1.6pc
\leftskip=1.6pc

\setcounter{footnote}{0}

$^\star$ Partly based on observations collected at the European Southern
Observatory, La~Silla, Chile

\f Osservatorio Astronomico di Bologna, Via Zamboni 33, 40126 Bologna, Italy

\f Departamento de Astronomia, Instituto de F\'\i sica, Universidade Federal do
Rio Grande do Sul, Avenida Bento Gon\c calves, 9500, CP15051, CEP 91500, Porto
Alegre, RS, Brasil

\f Visiting Astronomer at the Cerro Tololo Interamerican Observatory of the
National Optical Observatories, operated by AURA under contract with the
National Science Foundation

\f European Southern Observatory, Casilla 19001, Santiago 19, Chile

\f ISO Observatory, ESA Astrophysics Division, ESTEC, Postbus 299, 2200 AG
Noordwijk, The Netherlands

\f Observatoire de Paris, URA173 CNRS, Universit\'e Paris~7, Place Jules
Janssen, 92195, Meudon Principal Cedex, France

\f Observatorio Astronomico de Cordoba, Cordoba, Argentina

\f Visiting astronomer at the Complejo Astronomico EL Leoncito (CASLEO), San
Juan, Argentina

\f Inter-University Centre for Astronomy and Astrophysics, Post Bag 4,
Ganeshkhind, Pune 411007, India

\f Department of Astronomy, University of Cape Town, Private Bag, 7700
Rondebosch, Cape Town, South Africa

\f Department of Astronomy, The Ohio State University, 174 West 18th Avenue,
Columbus, OH 43210

\f Postal address: Lowell Observatory, Mars Hill Road, 1400 West, Flagstaff, AZ
86001

\f Royal Greenwich Observatory, Madingley Road, Cambridge CB3 0HA, United
Kingdom

\f Dipartimento di Astronomia, Universit\`a di Padova, Vicolo
dell'Osservatorio~5, 35122 Padova, Italy

\f Osservatorio Astronomico di Capodimonte, Via Moiariello 16, 80131 Napoli,
Italy

\f Vainu Bappu Observatory, Indian Institute of Astrophysics, Kavalur,
Alangayam 635701, Tamil Nadu, India

\f Observatorio Astronomico Felix Aguilar, San Juan, Argentina

\f Yale Southern Observatory, San Juan, Argentina

\f South African Astronomical Observatory, P.O. Box 9, Observatory 7935, Cape
Town, South Africa

\f Observatorio Nacional, CNPq, Rio de Janeiro, Brasil

\f Indian Institute of Astrophysics, Bangalore 560 034, India

\f Observatorio Astronomico de La Plata, La Plata, Argentina

\f Kiso Observatory, Institute of Astronomy, University of Tokio, Japan

\f Astronomisch Instituut, Katholieke Universiteit Leuven,
Celestijnenlaan~200~B, 3001 Heverlee, Belgium

\f Department of Physics, Vista University, Soweto Campus, Private Bag X09,
2013
Bertsham, Johannesburg, South Africa

\f Universit\"ats-Sternwarte, Geismarlandstra\ss e 11, W-3400 G\"ottingen,
Germany

\f Department of Physics and Astronomy, UCL, Gower Street, London WC1E 6BT,
United Kingdom

\f Istituto Astronomico dell'Universit\`a, Via Lancisi 29, 00161 Roma, Italy

\f ESA IUE Observatory, P.O. Box 50727, 28080 Madrid, Spain

}

\begin{center}
ABSTRACT
\end{center}

The Seyfert~1 galaxy NGC~3783 was intensely monitored in several bands
between 1991 December and 1992 August. This paper presents the results from the
ground-based observations in the optical and near-IR bands, which complement
the data-set formed by the International Ultraviolet Explorer (IUE) spectra,
discussed elsewhere. Spectroscopic and photometric data from several
observatories were combined in order to obtain well sampled light curves of the
continuum and of \hb. During the campaign the source underwent significant
variability. The light curves of the optical continuum and of \hb\ display
strong similarities with those obtained with the IUE. The near-IR flux
did not vary significantly except for a slight increase at the end of the
campaign.

The cross correlation analysis shows that the variations of the optical
continuum have a lag of 1~day or less with respect to those of the UV
continuum, with an uncertainty of $\le4$~days. The integrated flux of \hb\
varies with a delay of about 8~days. These results confirm that (a) the
continuum variations occur simultaneously or with a very small lag across the
entire UV-optical range, as in the Seyfert galaxy NGC 5548; and (b) the
emission lines of NGC 3783 respond to ionizing continuum variations with less
delay than those of NGC~5548. As observed in NGC~5548, the lag of \hb\ with
respect to the continuum is greater than those of the high ionization lines.

\bigskip

\noindent
{\it Subject headings:\/} galaxies: active --- galaxies: individual (NGC~3783)
--- galaxies: nuclei --- galaxies: Seyfert

\newpage
\begin{center}
1. INTRODUCTION
\end{center}

The nucleus of the Seyfert~1 galaxy NGC~3783
($\alpha,\delta_{1950.0}$=113633.0$-$372741, $z=0.0097$, $m_V\sim13$) is one of
the brightest and therefore best-studied active galactic nuclei (AGN) of the
Southern hemisphere. The optical spectrum of this source, which has been
extensively studied by Pelat, Alloin, \& Fosbury (1981) and Evans (1988),
presents very strong emission lines, with a high degree of ionization (Fig.~1).
Variations of the broad components of the optical emission lines have been
reported by Menzies \& Feast (1983), Stirpe, de~Bruyn, \& van~Groningen (1988),
Evans (1989), Winge, Pastoriza, \& Storchi-Bergmann (1990), Winge et al.\
(1992), and Winkler (1992). De~Ruiter \& Lub (1986) and Winkler et al.\ (1992)
present multi-band photometric observations, which show that the source has
varied by several tenths of magnitude. Photometric U data shown by Glass (1992)
display variations in excess of 1.1~mag. Variations have also been observed in
the near-IR by Glass (1992), from $\sim0.5$~mag at J (1.25~$\mu$m) to
$\sim1.1$~mag at L (3.5~$\mu$m).

Because of this history of variability, NGC~3783 was judged a suitable
candidate for an intensive monitoring campaign, similar to that carried out
on NGC~5548 by the International AGN Watch consortium (Clavel et al.\
1991, Peterson et al.\ 1991, 1992, Dietrich et al.\ 1993, and Peterson et al.\
1993, henceforth Papers~I--IV and VII respectively; Maoz et al.\ 1993; see also
the review by Peterson 1993). The monitoring of NGC 3783 took place between
December~1991 and August~1992, and made use of the International Ultraviolet
Explorer (IUE) and of several ground-based telescopes, mainly in the Southern
hemisphere. The results of the IUE observations are presented by Reichert et
al.\ (1993, henceforth Paper~V). This paper presents the first results from the
ground-based spectroscopic and photometric observations in the optical and
near-IR bands. Section~2 gives a description of the observations, Section~3
describes how the spectra were corrected for slit losses, Sections 4 and~5
describe how the light curves of the continuum and of \hb\ were obtained,
Section~6 presents a discussion and analysis of the light curves, and Section~7
summarizes our main conclusions.

\bigskip
\begin{center}
2. OBSERVATIONS\\
\medskip
{\it 2.1. Spectroscopy}
\end{center}

Spectroscopic observations were obtained at Cerro Tololo Interamerican
Observatory (CTIO) in Chile, at the European Southern Observatory (ESO), La
Silla, Chile, at the Complejo Astronomico El Leoncito (CASLEO), Argentina, at
the Ohio State University (OSU) Perkins reflector at Lowell Observatory, USA,
and at Vainu Bappu Observatory, Kavalur, India. All spectra were recorded on
CCDs, with the exception of those taken at the CTIO 1m telescope, where the
2D-Frutti detector was used. The bulk of the data comes from this telescope and
from the ESO 1.5m telescope: the CTIO spectra were all obtained with the same
instrumental set-up, and therefore form a very homogeneous data-set; the ESO
spectra vary widely in resolution and wavelength coverage, because the
program shared nights with other scheduled observers, who had priority in the
choice of the grating. A journal of the observations is given in Table~1: the
columns list respectively the Julian $\rm Date-2440000$ and UT date at the
midpoint of the integration, a code (A--E) which indicates where the spectrum
was obtained, the airmass at the midpoint of the integration, the area of the
aperture ($\rm slit\ width\times length$) in arcsec$^2$, the resolution and
wavelength range in \AA, and the integration times in seconds. The seeing was
generally between 1 and 2~arcsec. Epochs with sub-arcsecond seeing include
JD2448609, JD2448649, and JD2448824. Bad seeing ($\sim3$~arcsec) was recorded
on JD2448820.

All spectra were reduced using standard techniques. This paper presents the
light curve of the optical continuum and \hb\ only; the analysis of other
emission lines will be presented in a future paper. Unless otherwise
stated all wavelengths quoted in this paper are in the observed frame.

\medskip

\begin{center}
{\it 2.2. Optical photometry}
\end{center}

Photoelectric UBV(RI)$_C$ photometry was obtained at the 0.5m telescope of the
South African Astronomical Observatory at Sutherland, using the SAAO photometer
equipped with a Hamamatsu GaAs photomultiplier tube. An aperture of 20~arcsec
in diameter was used. The seeing varied between 1 and 5~arcsec. The
observations
and reduction were performed as described in Winkler et al.\ (1992). The
results are listed in Table~2. The uncertainties on the magnitudes are
0.03~mag.

Photoelectric UBV photometry was obtained at the 0.7m Charles D. Perrin
telescope at the Observatorio Astronomico Felix Aguilar (OAFA), San Juan,
Argentina, through an aperture of 33~arcsec in diameter. The photometer was
equipped with a RCA 31034A photomultiplier tube. The seeing was between 2 and
3~arcsec at all epochs. The data were calibrated using standard star fluxes
from Landolt (1973, 1983). The measurements are listed in Table~3. The first
part of the campaign was affected by the presence in the atmosphere of dust
from the Hudson volcano, which caused some color measurements to be unreliable:
these have been omitted from the table. The uncertainties on the magnitudes are
0.03~mag.

CCD photometry in BVR was obtained at the 1m telescope of the Vainu Bappu
Observatory (VBO). The seeing was between 2 and 3~arcsec for all observations.
Bias subtraction and flat fielding was performed on the frames, using Starlink
routines. The flux of NGC~3783 was measured with circular apertures with a
diameter of 18~arcsec, and calibrated with the close star SAO~202668, exposed
on the same frames. The results are listed in Table~4. The uncertainties are
0.05~mag for the magnitudes, and 0.03~mag for the colors.

There is a systematic difference between B$-$V values obtained at SAAO and VBO
on close dates, which cannot be accounted for entirely by the slightly
different
apertures used. Given the peculiar shape of AGN spectra, it is likely that the
difference is caused by the large color term required to convert the observed
VBO values to the Johnson photometric system. The accuracies quoted do not take
this systematic effect into account, but are correct within the individual
data-sets.
\medskip

\begin{center}
{\it 2.3. IR photometry}
\end{center}

JHKL photometry was obtained at SAAO, through a 12~arcsec aperture, using the
Mk~III infrared photometer attached to the 1.9m telescope. The observations and
reduction took place as described by Glass (1992). The chopping distance was
30~arcsec North and South of the nucleus. Tests with larger chopping distances
were made on NGC~3783 in March~1988: a distance of 60~arcsec yielded fluxes
which differed by at most 0.01~mag from those obtained with the 30~arcsec
distance. Therefore the latter was considered sufficient.

The fluxes were calibrated in the SAAO standard system (Carter 1990), using the
standard star HR4523. Other standards observed during the same nights provided
a check on the photometric accuracy. The magnitudes are given in Table~5. The
uncertainty of the J, H, and K magnitudes is 0.03 at all epochs. The
uncertainties of the L magnitudes are listed in the table.
\bigskip

\begin{center}
3. ABSOLUTE CALIBRATION OF THE SPECTRA
\end{center}

In order to correct for slit losses, the spectra which include \hb\ were
internally calibrated relative to each other with the method described by van
Groningen \& Wanders (1992). The method finds the optimum scaling factor,
wavelength shift, and convolution factor of one spectrum with respect to
another one, taken as reference, by slowly varying these parameters until the
residuals of one or more constant narrow lines in the difference between the
two spectra are minimized. One of the highest quality spectra (Fig.~1),
obtained at ESO on 1991 Dec.~18 (JD2448609), was used as reference, and the
residuals of the strong \oiiia\ and \la5007 lines were minimized. More than
90\% of the multiplicative scaling factors yielded by the scaling program are
between 0.8 and 1.5. The accuracy of the internal calibration is $\sim2\%$ for
most of the spectra, judging from the \oiii\ residuals obtained when slightly
varying the scaling factor. Although the ESO spectra were obtained through a
considerably narrower slit, differential light losses caused by the slightly
extended \oiii\ emission (Winge et al.\ 1992) were negligible: a comparison
between subsets of ESO and CTIO line fluxes obtained with separations of 2 days
at most showed that any systematic offset between the two data-sets is less
than 0.5\%.

The accuracy of the internal calibration method for spectra of very different
resolutions was tested by convolving some of the highest resolution spectra
with a Gaussian of appropriate width, in order to simulate the lowest
resolution used, and running the internal calibration program again: the
resulting scaling factors differed from those obtained for the non-convolved
spectra by 3\% at most, always in excess. This effect is partly compensated,
when measuring the line flux (see Section~5), by the fact that the lower
resolution causes a small fraction of the flux to be lost at the edges of the
integration interval: the resulting \hb\ flux, therefore, does not differ by
more than 2\%.

In order to obtain the absolute calibration of the spectra, the flux of the
\oiiib\ line was measured in the 5 CTIO spectra in which the \oiii\ lines were
strongest {\it before\/} the internal calibration: these spectra are of the
best quality and presumably suffered the least from slit losses or bad
observing conditions. The CTIO data set was chosen to obtain the absolute
calibration because of its homogeneity and large aperture. The  \oiiib\
measurements were obtained by fitting a straight pseudo-continuum under the
line and integrating the flux above it. The fluxes are listed in Table~6, with
their 1$\sigma$ uncertainties. The last line of the table gives the mean value
and standard deviation of the listed values. We henceforth assume that the flux
of \oiiib\ is $8.44\times10^{-13}$~\ergscm. The \oiiib\ flux measured in the
ESO spectrum used as reference for the internal calibration (JD2448609) is
$8.85\times10^{-13}$~\ergscm: therefore, all internally calibrated spectra were
multiplied by a factor 0.95 in order to obtain the right absolute calibration.

The value thus obtained for the \oiiib\ flux is about 15\% less than the flux
measured by Osmer, Smith, \& Weedman (1974), and we will therefore assume that
the systematic uncertainty of our absolute fluxes is of this order.
\bigskip

\begin{center}
4. THE CONTINUUM LIGHT CURVES\\
\medskip
{\it 4.1. Optical continuum}
\end{center}

The light curve of the optical continuum was obtained by combining spectral
with photometric data. The continuum measurements must not be significantly
contaminated by varying emission lines, which could introduce spurious delay in
the continuum variations: care was taken, therefore, to use a part of the
spectrum and a photometric band which are influenced as little as possible by
the broad lines. For the spectra, an additional constraint is given by the
necessity of obtaining the measurements in a region close to the \oiii\ lines,
used as internal calibrators, so that inaccuracies in the response calibration
do not introduce further uncertainties in the light curve.

The continuum flux was derived from the spectra by averaging the flux density
in an interval of 20~\AA\ centered at 5150~\AA, i.e.\ on the dip between the
\feiib\ and \la5169 lines which is considered to be the region close to \hb\
least contaminated by emission lines (Fig.~2). Although the two \feii\ lines
may have slightly overlapping wings in this interval, their variations have a
negligible influence on the continuum light curve.

The continuum measurements include a contribution from the underlying galaxy,
which has a different spatial distribution than the nucleus: the galaxy has an
apparent diameter of $\sim2$~arcmin, while the broad-line region (BLR) and the
region which emits the non-stellar continuum are unresolved. Therefore, only
the CTIO and OSU spectra were used, because of their similar large apertures: a
wide slit, in fact, minimizes the differential light losses caused by the
different spatial profiles of the nucleus and of the underlying galaxy, and
therefore ensures that the contribution of the stellar component to the
continuum fluxes depends only weakly on the seeing value. Because the seeing
was always lower than 2~arcsec during the CTIO observations, the stellar
component can be considered constant within the uncertainties.

The greatest source of uncertainty in the flux measurements is the internal
calibration error, which in all spectra of good quality is considerably larger
than the error introduced by the photon noise. Accordingly, most uncertainties
were set at 2\%. For some fluxes, obtained from spectra with lower
signal-to-noise ratio, the error bars were set at 4\%.

These points were then combined with the V fluxes, taken to represent
measurements of the optical continuum: this band was chosen not only because
its central wavelength is the closest to 5150~\AA, but also because it samples
a section of the optical spectrum which is almost free from strong variable
emission lines. The strongest broad lines are \heia\ and the \feii\ blend
centered at 5250~\AA\ (rest wavelength). The equivalent widths of these
features in the CTIO spectra are $<40$~\AA\ and $<25$~\AA\ respectively: we can
therefore assume that the variations of these lines, and their possible delays
with respect to those of the continuum, have a negligible effect on the V light
curve. The main narrow lines present in the V band are \feviia, which is weak,
and the \oiii\ lines: the latter introduce at most only a slight constant
offset in the V curve.

The V magnitudes were converted to flux densities using the relation
\begin{equation}
F_\lambda=10^{(-0.4V-8.47)}
\end{equation}
(Hayes \& Latham 1975), with $F_\lambda$ in \ergscma. Although the spectrum of
NGC~3783 does not have prominent emission features in the V band, this
conversion may nevertheless be affected by a systematic error because of the
differences between the spectrum of an AGN and that of the star (Vega) used to
define relation (1). Furthermore, as explained in Section~3, the spectroscopic
continuum fluxes have a systematic uncertainty of 15\%. For both these reasons
the continuum flux densities obtained from different data-sets were compared,
before being combined in a single optical continuum curve, as follows.

If the conversion of the photometric V data to flux densities is correct, and
if the absolute calibration of the spectroscopic fluxes is accurate, pairs of
quasi-simultaneous fluxes from one of the photometric data-sets and from the
spectroscopic data-set should align along a line of slope~1, and the
photometric fluxes should have a positive offset caused by the underlying
galaxy's higher contribution through the larger aperture. For each photometric
data-set, pairs formed by a photometric and a spectroscopic flux separated by
at most one day are plotted in Fig.~3. In each panel the continuous line is the
line of slope 1 which best fits the data. For the first two subsets (SAAO vs.\
spectroscopy and OAFA vs.\ spectroscopy), two least-squares linear regression
fits were calculated: ordinates against abscissae and viceversa. As shown in
Fig.~3, the slope of the bisector of the fits is close to one for both subsets.
Using the least-squares fit bisector to fine-tune the conversion would change
the photometric fluxes by 1$\sigma$ at most. The fits were not calculated for
the VBO vs.\ spectroscopy subset, as it consists of only three points. Figure~3
shows that the relation between these points is also well represented by a line
of slope~1.

This comparison of quasi-simultaneous pairs of points also allowed us to
correct
the photometric fluxes, in order to account for the different apertures used.
The underlying galaxy contributes in different amounts to each data-set. In
particular, the photometric measurements, obtained through apertures with
diameters of 18, 20, and 33~arcsec, are much higher than the spectroscopic
fluxes, obtained through a $5\times10$~arcsec$^2$ rectangular aperture. The
comparison between of quasi-simultaneous pairs of points yielded mean offsets
of $(8.62\pm0.74)\times10^{-15}$~\ergscma\ for the VBO fluxes,
$(9.71\pm0.66)\times10^{-15}$~\ergscma\ for the SAAO fluxes, and
$(16.64\pm0.51)\times10^{-15}$~\ergscma\ for the OAFA fluxes. These quantities
were subtracted from the photometric data, thus simulating photometry through a
$5\times10$~arcsec$^2$ rectangular aperture. The correction also automatically
eliminates the constant offset which may be caused by the presence of the
\oiii\ lines in the V band (see above). All fluxes are obviously still
contaminated by stellar light from the underlying galaxy: given the large
aperture, this component can be considered independent of seeing effects and
therefore constant. As discussed in detail by Alloin et al. (1993), examination
of the 2-D spectra allowed to estimate that the stellar contribution is about
30\% in the CTIO continuum measurements.

The continuum light curve obtained by combining the spectroscopic and
photometric data is listed in Table~7 and plotted in Fig.~4. Fluxes obtained
at the same date were averaged and their uncertainties averaged quadratically.
The median separation between consecutive points of the curve is 2.0~days.
The variations of the optical continuum strongly resemble those of the UV
continuum (Paper~V).

\medskip

\begin{center}
{\it 4.2. Near-IR continuum}
\end{center}

Figure 5 shows the IR light curves obtained at SAAO: the magnitudes were
converted to flux densities using the relations
\begin{eqnarray*}
F_\lambda(J) & = & 10^{(-0.4J-9.54)}  \\
F_\lambda(K) & = & 10^{(-0.4K-9.94)}  \\
F_\lambda(H) & = & 10^{(-0.4H-10.42)} \\
F_\lambda(L) & = & 10^{(-0.4L-11.16)}
\end{eqnarray*}
with $F_\lambda$ in \ergscma. The IR flux showed little evidence for variation
during the monitoring period and was generally lower than the long-term average
in the K (2.2~$\mu$m) and L (3.5~$\mu$m) bands. The only significant change was
an increase of $\sim20$\% at J, H, and K, and $\sim10$\% at L between JD2448768
and JD2448790. The wavelength trend of this variation was different from that
described by Glass (1992), who found that the amplitude of variation (in mag
and flux units) increases with wavelength in the JHKL bands.

There is no obvious similarity between the near-IR light curves and the UV and
optical ones. However, one should consider that the former are sampled more
sparsely, and that we expect them to be smoother than the UV and optical curves
because the region which emits the near-IR flux is likely to be much larger
than the BLR (Clavel, Wamsteker, \& Glass 1989; Glass 1992). Given the limited
amplitude and duration of the UV and optical variations, we cannot exclude that
corresponding features have simply been smoothed out of the near-IR light
curves.

There is no doubt as to the reality of the increase of flux around JD2448780,
but the spectral shape of the event is distinctly peculiar. The new H and K
flux observations have been plotted on Fig.~2b of Glass (1992) and three of
them (JD2888786, JD2448790, and JD2448804, which are also the brightest in J
and H) fall significantly away from the general trend, which is that H and K
are always linearly related. The reality of this interesting behaviour must at
present be considered provisional. A large number of observers were involved in
the work and small systematic differences between them cannot be entirely
excluded.

\bigskip

\begin{center}
5. THE LIGHT CURVE OF H$\beta$
\end{center}

The flux of \hb\ was measured as for NGC 5548 in Paper~II, by fitting a
straight-line continuum in the intervals 4800--4820~\AA\ and 5130--5150~\AA,
and integrating the flux above the line between 4830 and 4985~\AA\ (Fig.~2).
The constant narrow component of \hb\ is included in the measurements: its
contribution to the total line flux is $\sim7\times10^{-14}$~\ergscm\ (6--8\%
of the broad line flux, depending on the latter's strength).

Winge et al. (1992) have shown that the \ha\ emission extends to apparent
distances of $\sim10$~arcsec from the nucleus: there is therefore a possibility
that the \hb\ emission is also extended, and that seeing effects and different
slit widths may influence the \hb\ fluxes. The internal calibration is not
affected, because only the \oiii\ lines were used for its determination. The
broad line component (which forms the bulk of \hb) is emitted by an unresolved
region, which means that the broad line fluxes are not affected either. Any
extended \hb\ emission would only influence the narrow line component, causing
its flux to increase with increasing aperture and seeing. A strong effect is
unlikely, as the narrow component is only a small fraction of the measured \hb\
flux. The test described in Section~3 (the comparison of subsets of
quasi-simultaneous \hb\ fluxes from CTIO and ESO spectra) yields a negligible
offset between the two subsets. As the data-sets were obtained with different
apertures, we conclude that any extended narrow \hb\ emission adds a negligible
contribution to spectra obtained through large apertures.

The effect of a low resolution on the line flux measurement was simulated by
artificially degrading the spectra with the highest resolution, as described in
Section~3, and repeating the measurement: the \hb\ flux always decreased, but
never by more than 1\%. The 1$\sigma$ error bars were estimated as for the
continuum fluxes obtained from the spectra (Section~4).

The resulting fluxes and uncertainties are listed in Table~8. Again, fluxes
obtained during the same night were averaged. Figure~4 shows the \hb\ light
curve, in which the median separation between consecutive points is 2.1~days.
As for the optical continuum, the similarity with the IUE light curves
(Paper~V) is evident.
\bigskip

\begin{center}
6. ANALYSIS OF THE LIGHT CURVES
\end{center}

Inspection of the spectra already reveals that significant variability took
place in the optical band. As an example, Fig.~6 illustrates how the continuum
and broad lines underwent a strong decrease between JD2448661 and JD2448704.
Table~9 gives a summary of the main variability characteristics of the optical
and IR continua and of \hb, calculated as in Paper~V: the table lists the mean
value of each curve (each point was weighted with the inverse square of its
uncertainty), the reduced $\chi^2$ for variability with respect to the mean,
and the ratio between the maximum and minimum fluxes, R$_{max}$. The
variability is significant in all light curves except that of the L band flux.
The optical continuum varied with a lower amplitude than the UV continua
(Paper~V), and the multicolor photometry obtained at SAAO, OAFA, and VBO
(Tables 2--4) shows that the spectrum is harder when the source is more
luminous, confirming the trend displayed by the IUE data. This is due at least
in part to dilution by the stellar component. Stronger variability has been
observed in the past: the U flux varied by 0.38~mag at most during this
campaign, while differences in excess of 1.1~mag are present in the U light
curve shown by Glass (1992). The variation amplitude of \hb\ is lower than that
of most UV lines, and comparable to that of \civa. If a constant of
$7\times10^{-14}$~\ergscm\ is subtracted from all \hb\ fluxes, in order to
correct for the contribution of the narrow component, R$_{max}$ increases
slightly to $1.4\pm0.1$.

As already mentioned, a comparison between the light curves in Fig.~4 and those
presented in Paper~V reveals strong similarities. As well as the deep, broad
minimum observed in the UV continuum between JD2448680 and JD2448720, most of
the secondary minima also have counterparts in the optical curves: those at
JD2448631, JD2448668, and JD2448739 are visible in the continuum curve, and
those at JD2448668, JD2448728, JD2448739, JD2448767, and JD2448819 can be
observed, with a delay, in the light curve of \hb. Notice that in the \hb\
curve the minimum at JD2448767 (delayed to $\rm\sim JD2448775$) is of depth
comparable to that of the main minimum, while it is much less pronounced in the
UV continuum curves: the two minima have comparable depths also in the light
curve of \ciiia\ (Paper~V).

Unfortunately most of the period during which the IUE was observing at
intervals of 2~days instead of 4~days (JD2448783 to JD2448833) was disturbed by
bad weather at CTIO and ESO, leading to large gaps and poorer quality in the
optical light curves. Thus, while we cannot exclude that the short time-scale
`flickering' observed in the UV curves (Paper~V) is present also in the optical
light curves, the evidence for this is weak.

Figure 7 shows the cross correlation functions (CCF) of the optical continuum
and \hb\ light curves with the F$_{1460}$ continuum curve presented in Paper~V.
The CCFs were calculated with the method described by Gaskell \& Peterson
(1987), without artificially extending the light curves beyond their first and
last points (in other words, for each lag only the overlapping branches of the
curves were cross correlated). The discrete cross correlation functions (DCF)
were also calculated as described by Edelson \& Krolik (1988), with the
differences introduced in Paper~V. As Fig.~7 evidences, the CCF and DCF are in
good agreement. Notice that each CCF was obtained from light curves which were
derived from completely independent data-sets: there is therefore no influence
on the CCFs by correlated errors as when line and continuum fluxes are obtained
from the same spectra (Edelson \& Krolik 1988).

The window autocorrelation function (ACF) for the optical continuum (also shown
in Fig.~7) was obtained by averaging the ACFs of many white noise light curves
sampled with the same temporal pattern as the continuum curve (Gaskell \&
Peterson 1987). The full width at half maximum (FWHM) of the sampling window
ACF is 2.2~days, which indicates how much artificial correlation is introduced
by interpolating the light curve in order to calculate the CCF. The window ACF
for the \hb\ light curve is slightly broader ($\rm FWHM=2.7$~days). Both window
ACFs are much narrower than the CCFs, indicating that the variability on time
scales larger than about 2.5~days has been resolved.

The results of the CCF analysis are summarized in Table~10: for each CCF,
$\Delta t$(Peak) is the lag at which it reaches the maximum value, $\Delta
t$(Center) is the average of the two half-maximum lags, and $r_{max}$ is the
peak value. Cross correlating \hb\ and the optical continuum with any of the
three continuum light curves presented in Paper~V yields virtually identical
results. How to obtain a reliable uncertainty for the peak lag is still a
controversial matter: the method indicated by Gaskell \& Peterson (1987) gives
an uncertainty of 2~days for all the values of $\Delta t$(Peak) listed in
Table~10. The true uncertainty may be somewhat higher than this, but probably
does not exceed 4~days.

Notice that the FWHM of the CCFs are considerably larger than those of the CCFs
presented in Paper~V. One reason for this is the lower ratio between the
amplitude and noise of the variations in the optical light curves. Another
possible reason is their less regular sampling. In particular, if the CCFs are
calculated without including the points after JD2448777 (thus eliminating the
influence of the largest gap in the sampling pattern), the FWHM of the CCFs
become 22.8~days for the optical continuum and 35.3~days for \hb. Moreover, the
main minimum is broader in the \hb\ curve than in the continuum curves, and
this will naturally broaden the CCF of \hb. Paper~V presented evidence that the
variability of NGC~3783 may not be statistically stationary: the irregular
sampling of the ground-based campaign could therefore strongly influence the
shape of the CCF. The position of the CCF peak, however, is stable. The very
large FWHM of the \hb\ vs.\ F$_{2700}$ CCF is caused by the fact that its peak
is low enough to bring a broad plateau at positive lags just above the
half-maximum level.

The optical vs.\ UV continuum CCF reaches its maximum at 0 to 1~days: as for
NGC~5548 (Paper~II), therefore, there is no strong evidence that the optical
continuum is delayed with respect to the UV continuum. The lag between the
light curves of \hb\ and F$_{1460}$, as derived from the peak of the CCF, is
$\sim8$~days. This is a factor 2.5 less than the equivalent result for
NGC~5548, and confirms the difference in line vs.\ continuum lags between the
two objects, which was evidenced by the IUE data (Paper~V). Isolating the main
minimum in the optical and F$_{1460}$ curves and repeating the cross
correlation yields peak lags of 1 and 6~days for the optical continuum and \hb\
respectively. The last minimum observed in the \hb\ curve between JD2448820 and
JD2448844, when isolated, yields a lag of 8~days with respect to the
corresponding feature of the UV continuum. The difference between lags obtained
from isolated minima or from the entire light curves is not significant within
the uncertainties discussed above.

Because the near-IR light curves do not show any obvious similarity to the UV
and optical curves, we have not used them in the cross correlation analysis.
Glass (1992) found that the near-IR flux followed variations in the optical
continuum (U band) at an interval of 80 to 90 days. While the UV and optical
data do not exactly precede the IR by showing a step-like change at the
appropriate place, the dip around JD2448700 is followed by an increase
which precedes the IR step by about this amount. However, it is unlikely that
the two features are related: if they were, one would have to explain the lack
of a similar decrease in the IR flux before the brightening, and the fact that
the rise in flux occurs on comparable time scales (15--20~days) while the
emitting regions must be of considerably different sizes.
\bigskip

\begin{center}
7. SUMMARY AND CONCLUSIONS
\end{center}

We have presented the results obtained from the ground-based monitoring
of the Seyfert~1 galaxy NGC~3783, conducted in 1991--1992 in parallel
with the IUE monitoring campaign described in Paper~V. Light curves of the
optical continuum and of \hb, sampled at median intervals of 2.0 and 2.1~days
respectively, were obtained from spectroscopic and photometric observations
obtained at several observatories. The main results confirm the similarities
and differences between NGC~3783 and NGC~5548 evidenced by the IUE data, and
can be summarized as follows:

\begin{enumerate}

\item The optical continuum underwent significant variability, and its light
curve strongly resembles that of the UV continuum, with lower amplitude
variations.

\item Cross correlation analysis reveals no observable delay between the
variations of the optical and UV continua, within the (possibly conservative)
uncertainty of 4~days. This extends to the optical band the trend displayed by
the UV continuum light curves (Paper~V).

\item A significant increase occurred in the near-IR flux at the end of
campaign, but no feature in the UV/optical continuum curves is an obvious
`driver' for it. Considering the width of the K vs.\ U CCF in Glass (1992),
it is likely that the UV/optical continuum variations were not intense or long
enough for corresponding features, if any are present, to be visible in the
near-IR curves.

\item The light curve of \hb\ also strongly resembles those of the continua.
Cross correlation analysis reveals that its lag with respect to the UV
continuum is about 8~days. As in NGC~5548 (Papers I--III and VII), the light
curve of \hb\ in NGC~3783 is more delayed with respect to the continuum light
curve than those of the high ionization lines. The lag of \hb\ with respect to
the continuum is at least twice as long as that of \lya, about twice that of
\civ, and comparable to that of \mgiia.

\item The \hb\ vs. continuum lag in NGC~3783 is shorter than in NGC~5548, just
as the lags of the high ionization lines in NGC~3783 are shorter than their
counterparts in NGC~5548 (Paper~V).

\end{enumerate}

\bigskip
\bigskip

The authors are grateful to all the observatories involved for the generous
allocation of observing time, and to Drs.~J.~Baldwin, N.~Brosch, M.~Goad,
E.~J.~A.~Meurs, H.~Netzer, E.~P\'erez, E. Rokaki, J.~Roland, and W.~Wamsteker,
who supported the proposals but did not feel that their contribution was
sufficient for them to be co-authors. The Porto Alegre CCD camera, used for the
observations at CASLEO, is operated under a contract between UFRGS (Brasil) and
CASLEO (Argentina). CW acknowledges a Fellowship granted by the Brasilian
Institution CNPq. This work was partly supported by NSF grant AST-9117086 and
NASA grant NAG5-1824 (both to Ohio State University).

\newpage
\addtocounter{page}{5}
\begin{center}
TABLE 2\\
{\sc Photoelectric photometry from SAAO}\\
{\sc (20 arcsec aperture)}\\
\bigskip
\begin{tabular}{cccccc}
\hline
\hline
\\
JD & V & B$-$V & U$-$B & V$-$R$_C$ & V$-$I$_C$ \\
$-$2440000 \\
\\
\hline
\\
8610.56 & 12.93 & 0.63 & $-$0.67 & 0.53 & 1.00 \\
8622.55 & 12.96 & 0.61 & $-$0.65 & 0.56 & 1.02 \\
8630.57 & 12.96 & 0.66 & $-$0.60 & 0.55 & 1.04 \\
8634.55 & 12.98 & 0.65 & $-$0.60 & 0.56 & 1.03 \\
8638.57 & 12.96 & 0.65 & $-$0.63 & 0.57 & 1.04 \\
8644.56 & 12.94 & 0.60 & $-$0.65 & 0.56 & 1.02 \\
8651.58 & 12.94 & 0.63 & $-$0.66 & 0.55 & 1.02 \\
8655.58 & 12.93 & 0.60 & $-$0.68 & 0.55 & 1.03 \\
8659.54 & 12.92 & 0.60 & $-$0.68 & 0.55 & 1.01 \\
8663.53 & 12.93 & 0.62 & $-$0.66 & 0.57 & 1.04 \\
8671.52 & 12.98 & 0.61 & $-$0.59 & 0.56 & 1.03 \\
8684.52 & 13.00 & 0.67 & $-$0.60 & 0.57 & 1.05 \\
8688.60 & 13.11 & 0.68 & $-$0.58 & 0.58 & 1.07 \\
8700.45 & 13.05 & 0.72 & $-$0.55 & 0.58 & 1.06 \\
8704.42 & 13.08 & 0.70 & $-$0.57 & 0.58 & 1.09 \\
8714.41 & 13.03 & 0.63 & $-$0.65 & 0.55 & 1.02 \\
8717.50 & 12.99 & 0.64 & $-$0.63 & 0.54 & 1.03 \\
8729.38 & 12.97 & 0.63 & $-$0.65 & 0.56 & 1.04 \\
8733.36 & 12.97 & 0.63 & $-$0.64 & 0.56 & 1.02 \\
8743.40 & 12.96 & 0.65 & $-$0.63 & 0.55 & 1.03 \\
8747.34 & 12.96 & 0.63 & $-$0.64 & 0.55 & 1.04 \\
8802.25 & 12.96 & 0.60 & $-$0.65 & 0.56 & 1.03 \\
8803.27 & 12.92 & 0.61 & $-$0.66 & 0.54 & 1.00 \\
8823.22 & 12.91 & 0.62 & $-$0.65 & 0.51 & 1.02 \\
\\
\hline
\end{tabular}
\end{center}

\newpage
\begin{center}
TABLE 3\\
{\sc Photoelectric photometry from OAFA}\\
{\sc (33 arcsec aperture)}\\
\bigskip
\begin{tabular}{cccc}
\hline
\hline
\\
JD & V & B$-$V & U$-$B \\
$-$2440000 \\
\\
\hline
\\
8638.69 & 12.64 & \dots &   \dots \\
8653.73 & 12.67 & \dots &   \dots \\
8653.75 & 12.68 & \dots &   \dots \\
8661.75 & 12.65 & \dots &   \dots \\
8663.69 & 12.67 &  0.71 &   \dots \\
8665.76 & 12.69 &  0.72 & $-$0.70 \\
8665.77 & 12.68 &  0.70 & $-$0.60 \\
8690.65 & 12.72 &  0.71 & $-$0.50 \\
8716.66 & 12.68 &  0.68 & $-$0.54 \\
8743.54 & 12.70 &  0.69 & $-$0.48 \\
8809.54 & 12.61 &  0.66 & $-$0.61 \\
8833.52 & 12.60 &  0.63 & $-$0.56 \\
\\
\hline
\end{tabular}
\end{center}

\newpage
\begin{center}
TABLE 4\\
{\sc CCD photometry from VBO}\\
{\sc (18 arcsec aperture)}\\
\bigskip
\begin{tabular}{cccc}
\hline
\hline
\\
JD & V & B$-$V & V$-$R \\
$-$2440000 \\
\\
\hline
\\
8657.41 & 13.00 & 0.51 & 0.63 \\
8657.44 & 12.98 & 0.51 & 0.63 \\
8688.31 & 13.13 & 0.63 & 0.68 \\
8689.31 & 13.13 & 0.61 & 0.67 \\
8715.24 & 13.09 & 0.49 & 0.65 \\
\\
\hline
\end{tabular}
\end{center}

\newpage
\begin{center}
TABLE 5\\
{\sc Near-IR observations at SAAO}\\
{\sc (12 arcsec aperture)}\\
\bigskip
\begin{tabular}{ccccc}
\hline
\hline
\\
JD & J & H & K & L \\
$-$2440000 \\
\\
\hline
\\
8576 & 11.33 & 10.43 & 9.71 & 8.32$\pm$0.06  \\
8593 & 11.41 & 10.52 & 9.76 & 8.37$\pm$0.11  \\
8629 & 11.33 & 10.44 & 9.70 & 8.40$\pm$0.05  \\
8630 & 11.37 & 10.44 & 9.71 & 8.35$\pm$0.06  \\
8632 & 11.33 & 10.44 & 9.70 & 8.32$\pm$0.06  \\
8633 & 11.35 & 10.43 & 9.70 & 8.34$\pm$0.06  \\
8634 & 11.34 & 10.45 & 9.71 & 8.37$\pm$0.05  \\
8671 & 11.35 & 10.46 & 9.69 & 8.34$\pm$0.06  \\
8674 & 11.32 & 10.42 & 9.67 & 8.29$\pm$0.06  \\
8675 & 11.32 & 10.41 & 9.65 & 8.23$\pm$0.06  \\
8700 & 11.40 & 10.45 & 9.70 & 8.39$\pm$0.06  \\
8727 & 11.33 & 10.43 & 9.69 & 8.34$\pm$0.05  \\
8731 & 11.34 & 10.45 & 9.72 & 8.31$\pm$0.06  \\
8733 & 11.35 & 10.43 & 9.69 & 8.38$\pm$0.06  \\
8734 & 11.33 & 10.45 & 9.70 & 8.39$\pm$0.06  \\
8735 & 11.24 & 10.44 & 9.72 & 8.30$\pm$0.06  \\
8768 & 11.28 & 10.47 & 9.72 & 8.36$\pm$0.05  \\
8783 & 11.17 & 10.39 & 9.65 & 8.31$\pm$0.06  \\
8786 & 11.09 & 10.27 & 9.62 & 8.32$\pm$0.06  \\
8789 & 11.16 & 10.34 & 9.60 & 8.32$\pm$0.08  \\
8790 & 11.07 & 10.27 & 9.58 & 8.24$\pm$0.06  \\
8792 & 11.16 & 10.35 & 9.63 & 8.27$\pm$0.06  \\
8804 & 11.13 & 10.29 & 9.59 & 8.22$\pm$0.05  \\
8834 & 11.20 & 10.35 & 9.60 & 8.26$\pm$0.05  \\
\\
\hline
\end{tabular}
\end{center}

\newpage
\begin{center}
TABLE 6\\
{\sc Absolute flux of the \oiiib\ line}\\
\bigskip
\begin{tabular}{cc}
\hline
\hline
\\
JD & Flux \\
$-$2440000 & ($10^{-13}$ \ergscm) \\
\\
\hline
\\
8607 & 8.44$\pm$0.17 \\
8651 & 8.36$\pm$0.17 \\
8678 & 8.52$\pm$0.17 \\
8716 & 8.39$\pm$0.17 \\
8724 & 8.46$\pm$0.17 \\
\\
\hline
\\
Mean & 8.435$\pm$0.059
\end{tabular}
\end{center}

\newpage
\begin{center}
TABLE 7\\
{\sc Optical continuum light curve}\\
\bigskip
\begin{tabular}{ccc}
\hline
\hline
\\
JD & F$_\lambda$(opt) & Source \\
$-2440000$ & ($10^{-15}$ \ergscma) & \\
\\
\hline
\\
8607.83 & 13.02$\pm$0.26 & A \\
8610.56 & 14.19$\pm$0.63 & G \\
8622.55 & 12.57$\pm$0.62 & G \\
8623.83 & 12.90$\pm$0.26 & A \\
8627.83 & 12.35$\pm$0.25 & A \\
8630.57 & 12.57$\pm$0.62 & G \\
8631.84 & 12.30$\pm$0.25 & A \\
8634.55 & 12.16$\pm$0.61 & G \\
8635.83 & 11.89$\pm$0.24 & A \\
8638.63 & 12.79$\pm$0.50 & F,G \\
8639.84 & 12.67$\pm$0.25 & A \\
8643.72 & 14.07$\pm$0.28 & A \\
8644.56 & 12.98$\pm$0.63 & G \\
8647.76 & 13.91$\pm$0.28 & A \\
8651.67 & 12.75$\pm$0.26 & A,G \\
8653.74 & 12.30$\pm$0.57 & F \\
8655.58 & 13.19$\pm$0.63 & G \\
8656.77 & 13.61$\pm$0.27 & A \\
8657.42 & 13.04$\pm$0.71 & H \\
8659.54 & 13.40$\pm$0.64 & G \\
8660.77 & 13.24$\pm$0.26 & A \\
8661.75 & 13.15$\pm$0.82 & F \\
8663.53 & 12.96$\pm$0.50 & F,G \\
8664.77 & 12.37$\pm$0.25 & A \\
8665.77 & 12.03$\pm$0.56 & F \\
8668.81 & 11.31$\pm$0.23 & A \\
8671.52 & 12.16$\pm$0.61 & G \\
8672.72 & 13.44$\pm$0.54 & A \\
8676.71 & 11.97$\pm$0.24 & A \\
\\
\hline
\end{tabular}
\end{center}

\newpage
\begin{center}
TABLE 7---{\it Continued}\\
\bigskip
\begin{tabular}{ccc}
\hline
\hline
\\
JD & F$_\lambda$(opt) & Source \\
$-2440000$ & ($10^{-15}$ \ergscma) & \\
\\
\hline
\\
8677.80 & 12.28$\pm$0.25 & A \\
8678.68 & 12.09$\pm$0.24 & A \\
8684.52 & 11.76$\pm$0.59 & G \\
8685.80 & 10.12$\pm$0.20 & A$^\prime$ \\
8687.74 &  9.68$\pm$0.19 & A$^{\prime\prime}$ \\
8688.60 &  9.92$\pm$0.46 & G,H \\
8689.31 & 10.45$\pm$0.88 & H \\
8690.65 & 11.06$\pm$0.76 & F \\
8700.45 & 10.79$\pm$0.57 & G \\
8704.50 & 10.00$\pm$0.20 & A,G \\
8712.59 & 11.78$\pm$0.24 & A \\
8714.41 & 11.18$\pm$0.58 & G \\
8715.24 & 11.11$\pm$0.91 & H \\
8716.60 & 11.44$\pm$0.23 & A,F \\
8717.50 & 11.96$\pm$0.60 & G \\
8720.59 & 13.04$\pm$0.26 & A \\
8724.56 & 12.81$\pm$0.26 & A \\
8728.55 & 12.91$\pm$0.26 & A \\
8729.38 & 12.36$\pm$0.61 & G \\
8732.56 & 12.33$\pm$0.25 & A \\
8733.36 & 12.36$\pm$0.61 & G \\
8736.55 & 11.64$\pm$0.23 & A \\
8742.67 & 12.21$\pm$0.24 & B \\
8743.40 & 12.24$\pm$0.48 & F,G \\
8744.61 & 12.38$\pm$0.50 & A \\
8747.34 & 12.57$\pm$0.62 & G \\
8752.57 & 12.68$\pm$0.25 & A \\
8763.58 & 12.89$\pm$0.52 & A \\
8764.56 & 12.10$\pm$0.24 & A \\
\\
\hline
\end{tabular}
\end{center}

\newpage
\begin{center}
TABLE 7---{\it Continued}\\
\bigskip
\begin{tabular}{ccc}
\hline
\hline
\\
JD & F$_\lambda$(opt) & Source \\
$-2440000$ & ($10^{-15}$ \ergscma) & \\
\\
\hline
\\
8772.63 & 13.47$\pm$0.27 & A \\
8776.56 & 11.71$\pm$0.23 & A \\
8794.57 & 12.76$\pm$0.26 & A \\
8802.25 & 12.57$\pm$0.62 & G \\
8803.27 & 13.40$\pm$0.64 & G \\
8804.47 & 13.32$\pm$0.53 & A \\
8809.54 & 14.18$\pm$0.85 & F \\
8822.48 & 14.68$\pm$0.29 & A \\
8823.22 & 13.62$\pm$0.65 & G \\
8824.47 & 13.41$\pm$0.27 & A \\
8826.47 & 14.36$\pm$0.29 & A \\
8830.48 & 13.32$\pm$0.27 & A \\
8832.46 & 14.64$\pm$0.29 & A \\
8833.52 & 14.29$\pm$0.85 & F \\
\\
\hline
\end{tabular}
\end{center}
\bigskip
\bigskip
\begin{center}
\begin{tabular}{cl}
\multicolumn{2}{c}{Codes for data origin}\\
\\
\hline
\\
A & CTIO spectroscopy \\
B & OSU spectroscopy \\
F & OAFA photoelectric photometry \\
G & SAAO photoelectric photometry \\
H & VBO CCD photometry \\
\end{tabular}
\end{center}

\newpage
\begin{center}
TABLE 8\\
{\sc \hb\ light curve}\\
\bigskip
\begin{tabular}{ccc}
\hline
\hline
\\
JD & F(\hb) & Source \\
$-2440000$ & ($10^{-15}$ \ergscm) & \\
\\
\hline
\\
8593.83 & 1176$\pm$24 & D \\
8605.83 & 1123$\pm$22 & D \\
8607.83 & 1111$\pm$22 & A \\
8609.81 & 1118$\pm$22 & D \\
8613.84 & 1225$\pm$49 & D$^\prime$ \\
8621.83 & 1182$\pm$24 & D \\
8623.83 & 1111$\pm$22 & A \\
8627.83 & 1144$\pm$23 & A \\
8631.84 & 1183$\pm$24 & A \\
8633.81 & 1169$\pm$23 & C,D \\
8635.83 & 1130$\pm$23 & A \\
8637.78 & 1235$\pm$49 & C \\
8638.75 & 1199$\pm$24 & C \\
8639.84 & 1161$\pm$23 & A \\
8642.78 & 1125$\pm$23 & C \\
8643.72 & 1154$\pm$23 & A \\
8645.82 & 1174$\pm$23 & D \\
8647.76 & 1150$\pm$23 & A \\
8649.83 & 1183$\pm$24 & D \\
8651.76 & 1199$\pm$24 & A \\
8656.77 & 1198$\pm$24 & A \\
8657.86 & 1129$\pm$23 & D \\
8660.77 & 1156$\pm$23 & A \\
8661.85 & 1181$\pm$24 & D \\
8664.77 & 1227$\pm$25 & A \\
8668.81 & 1207$\pm$24 & A \\
8672.72 & 1067$\pm$43 & A \\
8676.71 & 1075$\pm$22 & A \\
8677.80 & 1071$\pm$21 & A \\
8678.68 & 1068$\pm$21 & A \\
\\
\hline
\end{tabular}
\end{center}

\newpage
\begin{center}
TABLE 8---{\it Continued}\\
\bigskip
\begin{tabular}{ccc}
\hline
\hline
\\
JD & F(\hb) & Source \\
$-2440000$ & ($10^{-15}$ \ergscm) & \\
\\
\hline
\\
8681.71 & 1117$\pm$22 & D \\
8685.79 & 1078$\pm$22 & A$^\prime$,D \\
8686.70 &  958$\pm$38 & C \\
8687.74 &  968$\pm$19 & A$^{\prime\prime}$ \\
8689.77 &  998$\pm$20 & C,D \\
8690.78 &  998$\pm$20 & C \\
8694.31 &  988$\pm$49 & E \\
8704.58 &  958$\pm$19 & A \\
8710.67 &  967$\pm$39 & D \\
8712.59 &  958$\pm$19 & A \\
8716.60 &  985$\pm$20 & A \\
8720.59 & 1036$\pm$21 & A \\
8724.56 & 1095$\pm$22 & A \\
8728.55 & 1179$\pm$24 & A \\
8732.56 & 1042$\pm$21 & A \\
8736.55 & 1068$\pm$21 & A \\
8737.55 & 1131$\pm$23 & D \\
8742.67 & 1112$\pm$22 & B \\
8744.61 & 1201$\pm$48 & A \\
8749.52 & 1039$\pm$21 & D \\
8752.57 & 1090$\pm$22 & A \\
8753.54 & 1063$\pm$21 & D \\
8764.56 & 1121$\pm$45 & A \\
8772.63 & 1106$\pm$22 & A \\
8776.56 & 1009$\pm$20 & A \\
8777.49 &  952$\pm$19 & D \\
8794.57 & 1182$\pm$24 & A \\
8800.52 & 1137$\pm$45 & D \\
8802.5  & 1198$\pm$48 & D \\
8803.5  & 1112$\pm$22 & D \\
\\
\hline
\end{tabular}
\end{center}

\newpage
\begin{center}
TABLE 8---{\it Continued}\\
\bigskip
\begin{tabular}{ccc}
\hline
\hline
\\
JD & F(\hb) & Source \\
$-2440000$ & ($10^{-15}$ \ergscm) & \\
\\
\hline
\\
8804.47 & 1273$\pm$51 & A \\
8805.5  & 1119$\pm$22 & D \\
8806.57 & 1194$\pm$48 & A \\
8808.47 & 1199$\pm$48 & A \\
8820.55 & 1258$\pm$25 & D \\
8822.51 & 1236$\pm$25 & A,D \\
8823.50 & 1193$\pm$24 & D \\
8824.50 & 1178$\pm$24 & A,D \\
8826.47 & 1131$\pm$23 & A \\
8830.47 & 1194$\pm$24 & A \\
8832.46 & 1203$\pm$24 & A \\
8841.49 & 1219$\pm$49 & D \\
8844.50 & 1239$\pm$50 & D \\
\\
\hline
\end{tabular}

\bigskip
\bigskip
Codes for origin of data as in Table 1
\end{center}

\newpage
\begin{center}
TABLE 9\\
{\sc Variability parameters}\\
\bigskip
\begin{tabular}{cccc}
\hline
\hline
\\
Feature & Mean flux$^1$ & $\chi^2_\nu\;^2$  & R$_{max}\;^3$ \\
\\
\hline
\\
F$_\lambda$(opt) & 12.4 & 12.0 & 1.5$\pm$0.1 \\
F$_\lambda$(J)   &  8.8 & 10.0 & 1.4$\pm$0.1 \\
F$_\lambda$(H)   &  7.8 &  4.2 & 1.3$\pm$0.1 \\
F$_\lambda$(K)   &  5.1 &  2.5 & 1.2$\pm$0.1 \\
F$_\lambda$(L)   &  3.4 &  0.8 & 1.2$\pm$0.1 \\
F(\hb)           & 1109 & 11.2 & 1.3$\pm$0.1 \\
\\
\hline
\\
\end{tabular}
\end{center}

\bigskip
\bigskip
\noindent $^1$ Weighted mean. Units are $10^{-15}$~\ergscma\ for continuum
light curves and $10^{-15}$~\ergscm\ for \hb\ light curve

\medskip

\noindent $^2$ Reduced $\chi^2$ for variability about mean. Degrees of freedom
are 71 for optical continuum, 23 for IR continua, and 72 for \hb

\medskip

\noindent $^3$ Ratio between maximum and minimum values. Uncertainties
calculated by propagation of 1$\sigma$ flux errors

\newpage
\begin{center}
TABLE 10\\
{\sc Cross correlation results}\\
\bigskip
\begin{tabular}{ccccc}
\hline
\hline
\\
Feature & $\Delta t$ (Peak) & $\Delta t$ (Center) & $r_{max}$ &  FWHM  \\
        &       (days)      &      (days)         &           & (days) \\
\\
\hline
\\
Opt.\ cont. vs.\ F$_{1460}$ & 1 &  1.6 & 0.741 & 41.9 \\
Opt.\ cont. vs.\ F$_{1835}$ & 0 &  1.3 & 0.746 & 42.2 \\
Opt.\ cont. vs.\ F$_{2700}$ & 0 &  2.1 & 0.760 & 44.4 \\
F(\hb) vs.\ F$_{1460}$      & 8 &  7.2 & 0.737 & 40.0 \\
F(\hb) vs.\ F$_{1835}$      & 8 &  7.4 & 0.742 & 39.9 \\
F(\hb) vs.\ F$_{2700}$      & 9 & 19.6 & 0.684 & 65.2 \\
\\
\hline
\end{tabular}
\end{center}

\newpage
\begin{center}
REFERENCES
\end{center}

\parindent=-1.6pc
\leftskip=1.6pc

Alloin D., et al.\ 1993, in preparation

Carter, B. S. 1990, MNRAS, 242, 1

Clavel, J., et al.\ 1991, ApJ, 366, 64 (Paper~I)

Clavel, J., Wamsteker, W., \& Glass, I. S. 1989, ApJ, 337, 236

de Ruiter, H. R., \& Lub, J. 1986, A\&AS, 63, 59

Dietrich, M., et al.\ 1993, ApJ, 408, 416 (Paper~IV)

Edelson, R. A., \& Krolik, J.H. 1988, ApJ, 333, 646

Evans, I. N. 1988, ApJS, 67, 373

Evans, I. N. 1989, ApJ, 338, 128

Gaskell, C. M., \& Peterson, B. M. 1987, ApJS, 65, 1

Glass, I. S. 1992, MNRAS, 256, 23P

Hayes, D. S., \& Latham, D. W. 1975, ApJ, 197, 593

Landolt, A.U., 1973, AJ, 78, 959

Landolt, A.U., 1983, AJ, 88, 439

Maoz, D., et al.\ 1993, ApJ, 404, 576

Menzies, J. W., \& Feast, M. W. 1983, MNRAS, 203, 1P

Osmer, P., Smith, M., and Weedman, D. W. 1974, ApJ, 189, 187

Pelat, D., Alloin, D., \& Fosbury, R. A. E. 1981, MNRAS, 195, 787

Peterson, B. M. 1993, PASP, 105, 247

Peterson, B. M., et al.\ 1991, ApJ, 368, 119 (Paper~II)

Peterson, B. M., et al.\ 1992, ApJ, 392, 470 (Paper~III)

Peterson, B. M., et al.\ 1993, ApJ, submitted (Paper~VII)

Reichert, G. A., et al.\ 1993, ApJ, in press (Paper~V)

Stirpe, G. M., de Bruyn, A. G., \& van Groningen, E. 1988, A\&A, 200, 9

van Groningen, E., \& Wanders, I. 1992, PASP, 104, 700

Winge, C., Pastoriza, M. G., \& Storchi-Bergmann, T. 1990, Rev.\ Mexicana
Astron.\ Astrof., 21, 177

Winge, C., Pastoriza, M. G., Storchi-Bergmann, T., \& Lipari, S. 1992, ApJ,
393, 98

Winkler, H. 1992, MNRAS, 257, 677

Winkler, H., Glass, I. S., van Wijk, F., Marang, F., Spencer Jones, J. H.,
Buckley, D. A. H., \& Sekiguchi, K. 1992, MNRAS, 257, 659

\newpage
\begin{center}
FIGURE CAPTIONS
\end{center}

Fig.\ 1: The spectrum of NGC 3783 obtained at ESO on 1991 December 18/19
(JD2448609). The wavelength range was covered entirely by a single exposure,
and is divided in two parts only for display purposes. The red part of the
spectrum is shown at two different scales in order to display \ha\ in its
entirety and to enhance the weaker lines. The scales in the lower panel differ
by a factor 4, and the vertical axis refers to the expanded scale. The letters
indicate the main features, as follows: a -- [O\,{\sc ii}]\,\la3727, b --
\fevii\,\la3760, c -- [Ne\,{\sc iii}]\,\la3869, d -- [Ne\,{\sc iii}]\,\la3968,
e -- H$\delta$, f -- H$\gamma$, g -- \oiii\,\la4363, h -- \feii\ multiplets, i
-- \heiia, j -- \hb, k -- \oiiia\ and \la5007, l -- \feviia, m -- \heia, n --
\feviib, o -- [O\,{\sc i}]\,\la6300, p -- [O\,{\sc i}]\,\la6363 and [Fe\,{\sc
x}]\,\la6375, q -- \ha, r -- [N\,{\sc ii}]\,\la6584, s -- [S\,{\sc
ii}]\,\la6717 and \la6731, t -- O$_2$ B-band (telluric), u -- \hei\,\la7065, v
-- [A\,{\sc iii}]\,\la7136.

\medskip

Fig.\ 2: The \hb\ region of the spectrum obtained at CTIO on 1992 February
11/12 (JD2448664). The straight lines illustrate the method used to measure the
fluxes of \hb\ and of the continuum: the continuous line indicates the
straight-line fit which approximates the continuum under \hb, the dashed lines
indicate the limits of the line flux integration above the continuum fit, and
the dotted lines indicate the limits of the interval used to average the
continuum flux.

\medskip

Fig.\ 3: Comparison between quasi-simultaneous subsets of continuum fluxes
obtained from the spectroscopic (CTIO) and photometric (SAAO, OAFA, and VBO)
data-sets. Units are $10^{-15}$~\ergscma. For each plotted point the
corresponding observations were separated in time by one day at most. The
photometric flux densities were obtained from the V data using equation (1).
The solid lines have slope~1 and intercept the y-axis at the mean difference
between ordinates and abscissae. This offset between different data-sets is
caused by the larger apertures used for the photometry, and therefore to a
larger flux from the underlying galaxy. The dashed lines represent the
bisectors of the least-squares linear regression fits calculated for ordinates
against abscissae and viceversa.

\medskip

Fig.\ 4: The light curves of the optical continuum (top panel) and of \hb\
(lower panel). Dotted vertical lines have been drawn to facilitate comparison
with Figs.~8--10 of Paper~V.

\medskip

Fig.\ 5: The near-IR fluxes obtained at SAAO.

\medskip

Fig.\ 6: Top: the spectra obtained at CTIO on JD2448660 (higher) and JD2448704
(lower). Bottom: difference between the two spectra shown in the top panel.
Notice how the continuum and all the main broad emission lines (particularly
\heiia) underwent a strong decrease.

\medskip

Fig.\ 7: Top: the CCF (continuous line) and DCF (full circles) of the
F$_{1460}$ continuum curve (Paper~V) versus the optical continuum curve shown
in Fig.~4. The DCF is calculated in 4-day bins. Middle: the CCF and DCF of the
F$_{1460}$ continuum curve versus the \hb\ curve shown in Fig.~4. Bottom: the
window ACF for the sampling pattern of the optical continuum. The window ACF
for the sampling of the \hb\ curve is slightly broader ($\rm FWHM=2.7$~days
instead of 2.1~days).

\end{document}